\documentclass{article}
\usepackage[utf8]{inputenc}
\usepackage{amssymb}
\usepackage{amsfonts}
\usepackage{mathrsfs}
\usepackage{amsmath}
\usepackage{appendix}
\usepackage{graphicx}
\usepackage{dsfont}
\usepackage{float}
\usepackage{diagbox}
\usepackage{subfigure}
\usepackage{natbib}
\usepackage{url}
\usepackage{graphics}
\usepackage{csquotes}
\usepackage[all]{xy}
\usepackage[margin=1.4in,marginparwidth=1in,marginparsep=.05in]{geometry}
\PassOptionsToPackage{hyphens}{url}
\usepackage{tikz-cd}
\usepackage{braket}
\usepackage{xcolor}
\usepackage{verbatim}
\usepackage{float}
\usepackage{mathrsfs}
\usepackage{extarrows}
\usepackage{tikz}
\usetikzlibrary{shapes,arrows,positioning}
\usepackage{mathrsfs}
\usepackage{mathtools}
\usepackage{authblk}

\usepackage{hyperref}
\usepackage{wrapfig}

\newcommand{\bF}{\mathbf F}

\newcommand{\bG}{\mathbf G}

\newcommand{\bS}{\mathbf S}

\newcommand{\bx}{\mathbf x}
\newcommand{\bT}{\mathbf T}
\newcommand{\bu}{\mathbf u}

\begin{document}

\title{\textbf{Leveraging Google's Tensor Processing Units for tsunami-risk mitigation planning in the Pacific Northwest and beyond}}
\author[1]{Ian Madden}
\author[2]{Simone Marras}
\author[1,3]{Jenny Suckale}

\affil[1]{Institute for Computational and Mathematical Engineering, Stanford University}
\affil[2]{Department of Mechanical Engineering \& Center for Applied Mathematics and Statistics, New Jersey Institute of Technology}
\affil[3]{Department of Geophysics, Doerr School of Sustainability, Stanford University}
\date{\today}



\maketitle

\begin{abstract}
Tsunami-risk and flood-risk mitigation planning has particular importance for communities like those of the Pacific Northwest, where coastlines are extremely dynamic and a seismically-active subduction zone looms large. The challenge does not stop here for risk managers: mitigation options have multiplied since communities have realized the viability and benefits of nature-based solutions. To identify suitable mitigation options for their community, risk managers need the ability to rapidly evaluate several different options through fast and accessible tsunami models, but may lack high-performance computing infrastructure. The goal of this work is to leverage the newly developed Google's Tensor Processing Unit (TPU), a high-performance hardware accessible via the Google Cloud framework, to enable the rapid evaluation of different tsunami-risk mitigation strategies available to all communities. We establish a starting point through a numerical solver of the nonlinear shallow-water equations that uses a fifth-order Weighted Essentially Non-Oscillatory method with the Lax-Friedrichs flux splitting, and a Total Variation Diminishing third-order Runge-Kutta method for time discretization. We verify numerical solutions through several analytical solutions and benchmarks, reproduce several findings about one particular tsunami-risk mitigation strategy, and model tsunami runup at Crescent City, California whose topography comes from a high-resolution Digital Elevation Model. The direct measurements of the simulations performance, energy usage, and ease of execution show that our code could be a first step towards a community-based, user-friendly virtual laboratory that can be run by a minimally trained user on the cloud thanks to the ease of use of the Google Cloud Platform.
\end{abstract}


\section{Introduction}
\begin{figure}[h]
    \centering
    \includegraphics[width=0.66\textwidth]{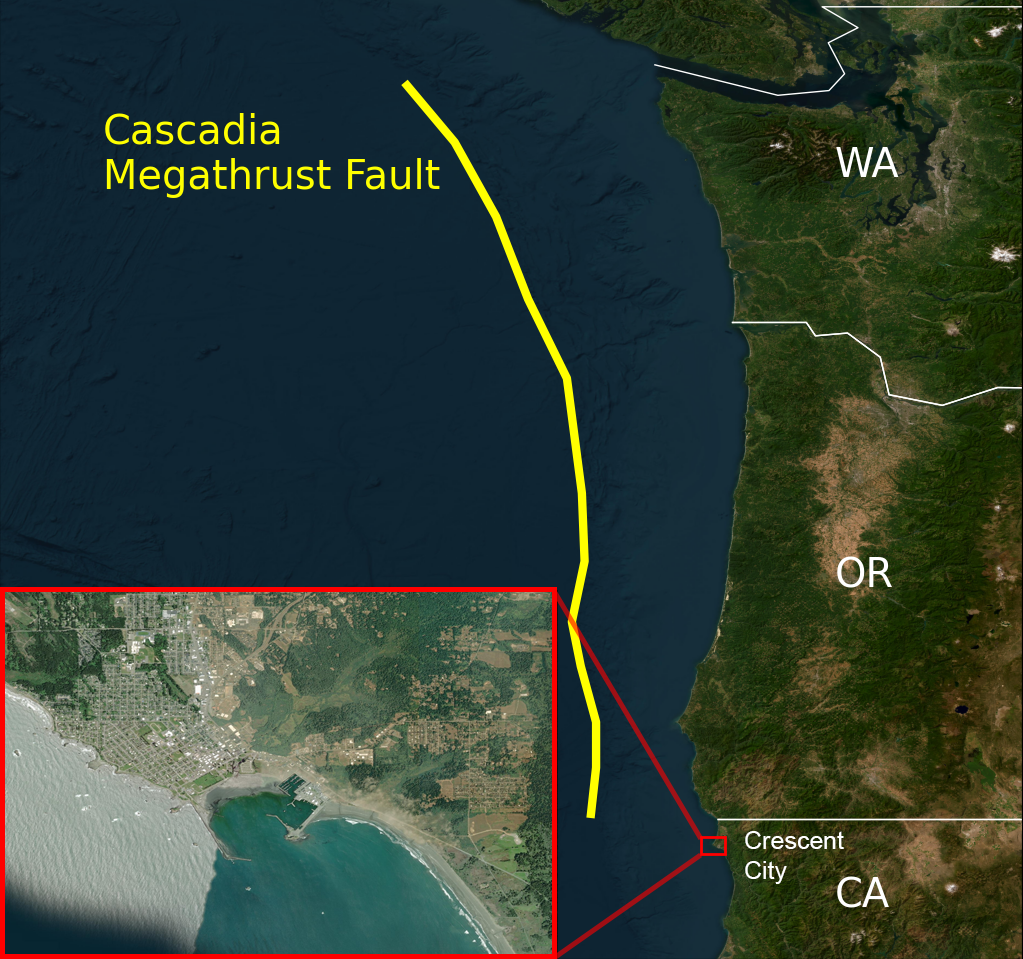}
    \caption{Map of the Cascadia Subduction Zone in the Pacific Northwest of the United States. Relative location of Crescent City with respect to the Megathrust Fault Line, with a more detailed picture of the Crescent City coastline. Esri provided access to the satellite imagery. Crescent City map at high resolution provided by Maxar. Pacific Northwest Map provided by Earthstar Geographics.}
    \label{fig:crescentintro}
\end{figure}

The coast of the Pacific Northwest, from Cape Mendocino in California to Northern Vancouver Island in Canada as depicted in Fig.~\ref{fig:crescentintro}, is located on the seismically active Cascadia subduction zone \citep{heaton1987earthquake,petersen2002simulations}. Along the 1200-km-long Cascadia subduction zone, there have been no large, shallow subduction earthquakes over the approximately 200 years of modern-data monitoring, but large historic earthquakes have left an unambiguous imprint on the coastal stratigraphy  \citep{clague1997evidence}. Sudden land level change in tidal marshes and low-lying forests provide testimony of 12 earthquakes over the last 6700 years \citep{witter2003great}, including one megathrust event that ruptured the entirety of the current Cascadia subduction zone in 1700 BC \citep{nelson1995radiocarbon,wang2013heterogeneous}. The event created a massive tsunami that swept across the entire Pacific Ocean devastating communities as far away as Japan \citep{satake1996time,atwater2011orphan}. Current seismic-hazard models estimate that the probability of another magnitude 9+ earthquake happening within the next 50 years is about 14\% \citep{petersen2002simulations}. 

A magnitude 9+ Cascadia earthquake and tsunami occurring during modern times would devastate many low-lying communities along the Pacific Northwest. A recent assessment suggests that deaths and injuries could exceed tens of thousands and entails economic damages in the order of several billions of dollars for Washington and Oregon State \citep[see, e.g.,][]{oregonReport2013cascadia,nisacTsunami}, with potentially severe repercussions for the entire Pacific coast and country as a whole. The tsunami itself would put tens of thousands at risk of inundation, and threaten the low-lying coastal communities specifically in the Pacific Northwest with very little warning time for evacuation \cite{nisacTsunami}. But how to confront this risk? Traditionally, the most common approach to reducing tsunami risk is the construction of sea walls, but this hardening of the shoreline comes at a staggering price in terms of the economic construction costs (e.g., 245 miles of sea walls in Japan cost \$12.74 billion) and in terms of long-term negative impact on coastal ecosystems \citep{petersonLowe2009, duganHubbard2010, bulleriChapman2010} and shoreline stability \citep{deanDalrymple2002, komar1998}.

\begin{figure}[h]
    \centering
    \includegraphics[width=0.5\textwidth]{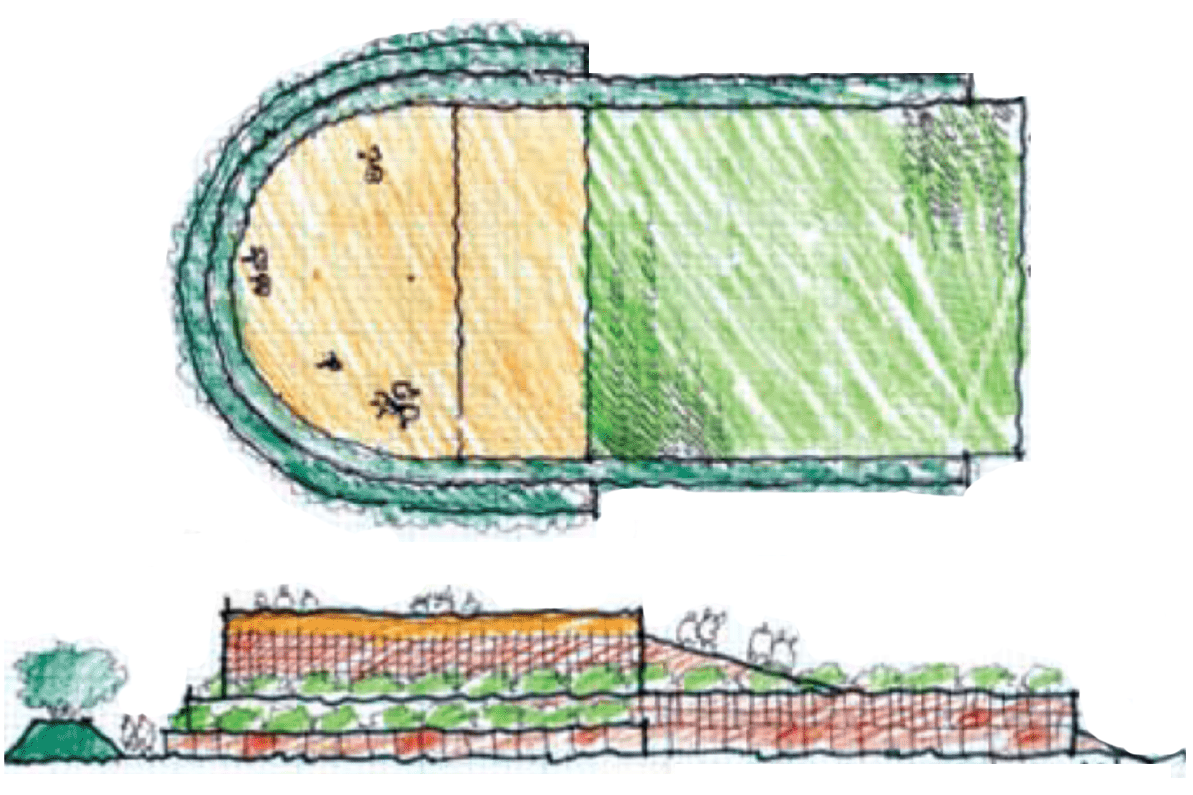}
    \caption{Map view (top) and side view (bottom) of a proposed tsunami-mitigation berm as designed by Project Safe Havens. The berm provides vertical evacuation space for the adjacent community and could also lower the onshore energy flux that drives the damage created by tsunami impact. We show this design as one example of a hybrid approach to tsunami-risk mitigation as it combines an engineered hill and ramp with natural vegetation. Sketches adapted from \citet{safeHavenV1}.}
    \label{fig:coastalmitigationpark}
\end{figure}

A potentially appealing alternative to sea walls are so-called hybrid approaches. Hybrid risk mitigation combines nature-based elements and traditional engineering elements to reduce risk while also providing co-benefits to communities and ecosystems. An example of a hybrid approach to tsunami-risk mitigation is a coastal mitigation park: A landscape unit on the shoreline built specifically to protect communities or critical infrastructure and provide vertical evacuation space, in the styles of Fig.~\ref{fig:coastalmitigationpark}. Communities across the Pacific Northwest are increasingly considering these nature-based or hybrid options \citep{safeHavenV1}, but many important science questions regarding protective benefits and optimal design remain open \citep{lunghinoEtAl2019, mukherjeeEtAl2023}. This gap is particularly concerning given that existing models show that a careful design is necessary to avoid potential adverse effects \citep{lunghinoEtAl2019}. The design of current mitigation parks, such as the one being built in Constituci\'on, Chile, is not yet underpinned by an in-depth quantification of how different design choices affect risk-reduction benefits, partly because numerical simulations of tsunami impacts are computationally expensive.

The goal of this paper is to leverage Google's Tensor Process Units (TPUs) for enabling a fast evaluation of different mitigation park design and ultimately advancing evidence-based tsunami-risk mitigation planning rooted in quantitative assessments. TPUs are a new class of hardware accelerators developed by Google with the primary objective of accelerating machine learning computations. They are accessible via the Google Cloud Platform \citep{jouppiEtAl2017, googleTPU} and have recently been used for many different applications in computational physics and numerical analysis \citep{pierceEtAl2022, luEtAl2020a, luEtAl2020b, bellettiEtal2020}. We build upon and extend an existing implementation \citep{pierceEtAl2022} to simulate the impact of idealized tsunamis on the coastline. Our implementation of the shallow water equations includes a non-linear advective term, not considered in \citep{pierceEtAl2022}, within the software framework based on Google's {\tt TensorFlow} necessary to execute code on the TPU. We discretize the new equations using the Weighted Essentially Non-Ocillatory (WENO) method \citep{weno1994} and a third-order Runge-Kutta in time. 

Numerical simulations of tsunamis have contributed to our understanding of and ability to mitigate wave impacts for many decades now, starting from the early work by \citet{isozaki1964numerical} for Tokyo Bay, and \citet{ueno1960numerical} for the Chilean coast. The ability to capture the rupture mechanism that generates the initial condition for tsunami propagation then enabled the reproduction of many historical tsunamis \citep{aida1969, aida1974}. Since then, many numerical models have been developed to simulate tsunami generation \citep{borreroEtAl2004, peltiesEtAl2012source, lopezEtAl2014tsunami, galvezEtAl2014, ulrichEtAl2019}, propagation \citep{titovEtAl2005, leveque2011tsunami, chenEtAl2014, allgeyerCummins2014, abdolaliKirby2017, bonev2018discontinuous, abdolaliEtAl2019}, and inundation \citep{lynett2007effect, parkEtAl2013, leschkaEtAl2014, chenEtAl2014, marsooliWu2014, mazaEtAl2015, oishiEtAl2015, prasetyoEtAl2019, lunghinoEtAl2019} by solving different variations of the shallow water and Navier-Stokes equations. 

The list of existing numerical models is long and was recently reviewed in \citet{marrasMandli2021} and \citet{horrilloEtAl2015}. Some commonly used ones are FUNWAVE \citep{kennedyEtAl2000funwave,shiEtAl2012funwave}, pCOULWAVE \citep{lynettEtAl2002COULWAVE, kimLynett2011COULWAVE}, Delft3D \citep{delft3D}, GeoCLAW \citep{berger2011geoclaw}, NHWAVE \citep{maEtAlNHWAVE2012}, Tsunami-HySEA \citep{maciasEtAl2017,maciaEtAl2020field,maciasEtAl2020lab}, FVCOM \citep{chenEtAl2003,chenEtAl2014}. Our work here relies on well-known numerical techniques to solve idealized tsunami problems. Its novelty lies in demonstrating the capability and efficiency of TPUs to solve the non-linear shallow water equations to model tsunamis. 

We intentionally use a hardware infrastructure that is relatively easy to use without specific training in high-performance computing (See the Google Cloud TPU page at \citet{googleTPU}), and may become a standard hardware on which physics-based machine-learning algorithms will be built \citep{raspEtAl2018, maoEtAl2020, wesselsEtAl2020, fauziMizutani2020, liuEtAl2021ML, kamiyaEtAl2022}. We propose that our implementation is one step towards a community-based, user-friendly virtual laboratory that can be run by a minimally trained user on the cloud thanks to the ease of use of the Google Cloud Platform. The tool, which is freely available on Github at \citep{tsunamiTPUlab} under an Apache License, Version 2.0 for collaborative open source software development, can be modified to include machine learning capabilities and, eventually, extended to coupled models for earth-quake generation, inundation, and human interaction.

\section*{Methods}
\subsection*{Numerical approximation}

We model tsunami propagation and runup with the 2D non-linear shallow water equations in the conservative formulation with a source term in a Cartesian coordinate system. Letting $\bx=(x,y)$ denote position, we define $u(\bx,t)$ and $v(\bx,t)$ as the flow velocities in the $x$ and $y$ directions, respectively. In our implementation, we solve for $h$, $hu$, and $hv$. We define $h(\bx,t)$ as the dynamic water height and $b(\bx)$ as the imposed bathymetry, meaning that the quantity $h + b$ represents the water surface level. This leads to the following system of equations, a set very similar to that suggested by \citet{Xing2005}
\begin{eqnarray} 
\label{eq:continuity} &\frac{\partial}{\partial t}h + \frac{\partial}{\partial x}(hu) + \frac{\partial}{\partial y}(hv) = 0 \\
\label{eq:x_momentum}&\frac{\partial}{\partial t}(hu)  + \frac{\partial}{\partial x}\left(\frac{(hu)^2}{h} + \frac{1}{2}g(h^2-b^2) \right) + \frac{\partial}{\partial y}(huv) = -g(h+b)\frac{\partial b}{\partial x} - \frac{gn^2\sqrt{(hu)^2+(hv)^2}}{h^{7/3}}(hu)  \\
\label{eq:y_momentum}&\frac{\partial}{\partial t}(hv)  + \frac{\partial}{\partial x}(huv) + \frac{\partial}{\partial y}\left(\frac{(hv)^2}{h} + \frac{1}{2}g(h^2-b^2) \right)  = -g(h+b)\frac{\partial b}{\partial x}  - \frac{gn^2\sqrt{(hu)^2+(hv)^2}}{h^{7/3}}(hv) \, ,
\end{eqnarray}
where $g = 9.81$ ms$^{-2}$ is the acceleration of gravity, and $n$ is the Manning friction coefficient.

For ease of future notation, we let $\bu = \begin{bmatrix} h & hu & hv \end{bmatrix}^T$, and we rewrite the above equations in a vector form, namely:
\begin{equation}
\label{eq:vector-form}
\frac{\partial \bu}{\partial t} + \frac{\partial \bF}{\partial x} + \frac{\partial \bG}{\partial y} = \bS
\end{equation}
where $\bF$ and $\bG$ are the fluxes in the $x$ and the $y$
directions for the vector $\bu$, and $\bS$ is a source term arising from variations in topography and Manning coefficient.

We implement these shallow-water equations using the finite volume method whereby the half-step flux and height values are determined through a 5th-order WENO scheme \citep{weno1994,Jiang1996}. We approximate solutions to cell-wise Riemann problems by formulating fluxes using the Lax-Friedrichs method as in  \citep{leveque_2011}. We formulate the bed source term as suggested by \citet{Xing2005}, and formulate the friction term explicitly rather than using the implicit process suggested by \citet{Xia2018}. We use a 3$^{rd}$-order, Total Variation Diminishing Runge-Kutta scheme to step the numerical solution forward in time.

We begin the discretization of the equation in continuous variables $t$, $x$, and $y$, using respective step sizes $\Delta t$, $\Delta x$, and $\Delta y$, which indicate the distance between consecutive integral steps in the discrete indices $n$, $i$, and $j$, respectively. We use the 5th-order WENO scheme in the $x$ and $y$ directions, where two values of each quantity $h$, $hu$, and $hv$ are determined at each half-step of $x$ and $y$. These two values correspond to a positive and negative characteristic, due to the nature of the footprint that is chosen at a given point. In other words, given the conservative form with relevant variable $\bu$, $\bu_{i,j}$ centered on a finite volume cell, we label these outputs of WENO:
\begin{equation}
\bu_{i+\frac{1}{2},j}^+ , \bu_{i+\frac{1}{2},j}^- \quad \text{for WENO in $x$, \quad or} \quad \bu_{i,j+\frac{1}{2}}^+ , \bu_{i,j+\frac{1}{2}}^- \text{for WENO in $y$.}
\end{equation}
From here, we use the Lax-Friedrichs method to approximate flux values that serve as solutions to the Riemann problem; i.e. we approximate
\begin{equation}
\bF_{i+\frac{1}{2},j} = \frac{1}{2} \left [ \bF( \bu_{i+\frac{1}{2},j}^+) + \bF(\bu_{i+\frac{1}{2},j}^-) - \alpha_\bu \left ( \bF( \bu_{i+\frac{1}{2},j}^+) - \bF(\bu_{i+\frac{1}{2},j}^- \right ) \right ]
\end{equation}
where $\alpha_\bu$ is the associated Lax-Frierichs global maximum characteristic speed. Now, we discretize Eq.~\ref{eq:vector-form} explicitly as:
\begin{equation}
\label{eq:discretized-conservation}
\frac{\bu^{n+1}_{i,j} - \bu^n_{i,j}}{\Delta t} + \frac{\bF^{n}_{i+\frac{1}{2},j} - \bF^{n}_{i-\frac{1}{2},j}}{\Delta x} + \frac{\bG^{n}_{i,j+\frac{1}{2}} - \bG^{n}_{i,j-\frac{1}{2}}}{\Delta y} = \bS(\bu^n_{i,j})
\end{equation}
Note that in our case, we also choose to formulate the source term $\bS(\bu^n_{i,j})$ explicitly and centered at the grid point. Since we use an entirely explicit formulation, we can rewrite Eq.~\ref{eq:discretized-conservation} as a time stepping operator for $\bu^{n+1}$, namely $\bu^{n+1} = \bT(\bu^{n})$. Because Runge-Kutta uses multiple stages within each time step, we reassign the output of the $\bT$ operator to be $\bu^{(n+1)} = \bT(\bu^{n})$, where$(n+1)$ indicates an intermediate full time step forward. This means a full Runge-Kutta step progresses as follows:
\begin{equation}\label{eq:RK-process} 
\bu^{(n+2)} = \bT(\bT(\bu^{n})) \quad \longrightarrow 
 \quad \bu^{(n+\frac{3}{2})} = \bT(0.25\bu^{(n+2)}+0.75\bu^{n}) \quad \longrightarrow \quad \bu^{n+1} = \frac{2}{3}\bu^{(n+\frac{3}{2})} + \frac{1}{3}\bu^n
\end{equation}
The process outlined by Eq.~\ref{eq:RK-process} outputs a final $\bu^{n+1}$ representing a full-step forward in simulation time.
\subsection*{TPU implementation}
\begin{figure}[h]
    \centering
    \includegraphics[width=\textwidth]{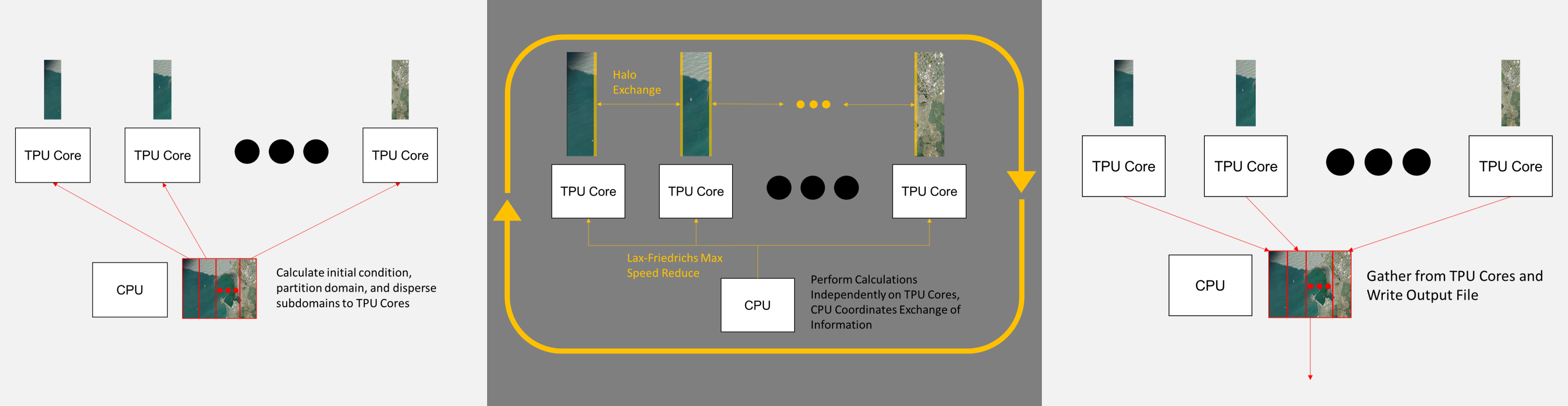}
    \caption{Left: Initialization of implementation takes advantage of CPU to allocate initial conditions and topography. Center: Regular computation period occurring on each subdomain, run independently on TPU cores with some data sharing coordinated by CPU. Right: CPU Gather to write results to output files.}
    \label{fig:tpuimplementation}
\end{figure}

To leverage the TPU's several cores, we divide the domain into multiple subdomains and independently compute the numerical solution to the governing equations on each core. While a lot of the computation can take place independently, each subdomain remains dependent on the others via their boundaries and the Lax-Friedrichs global maximum in characteristic speed. We determine global maximum characteristic speed by sharing and reducing the Lax-Friedrich maximum characteristic speed calculated on each core. We transfer subdomain boundary information with further care by using a halo exchange. The data transfer behavior and computation structure is summarized in Fig.~\ref{fig:tpuimplementation}.

Our implementation is inspired by \citet{pierceEtAl2022}, who chose halo exchange as an instrument for the TPU to communicate information across subdomain boundaries in their formulation of the shallow water equations. In the halo exchange process, we transfer slices of the domain from one core to the others immediately adjacent. While \citet{pierceEtAl2022}'s methodology only involved the exchange of a single slice from one core to the other, we transfer several slices in order to take full advantage of the high accuracy and larger footprint of the WENO scheme. These halo exchanges are then performed in every stage of the Runge-Kutta scheme, meaning that they occur multiple times in a single time step. 

The initial conditions and results are communicated from the remote program, which resides on the CPU, to the TPU workers by means of {\tt tpu.replicate} which sends {\tt TensorFlow} code to each TPU. We refer to \citet{pierceEtAl2022} for further details on the TPU implementation.

\section{Model verification and validation}
%
We differentiate between model verification and validation in the manner suggested by \citet{Carson}. Specifically, we check for model and implementation error by quantifying the extent to which numerical solutions compare to correct analytical solutions \citep{Carson}: wet dam break (Section 2.1), oscillations in a parabolic bowl (Section 2.2), and steady state flow down a slope with friction (Section 2.3). Following this, we validate by checking how well numerical solutions reflect the real system and apply to the context \citep{Carson}. To do this, we compare against an existing numerical benchmark from the Inundation Science and Engineering Cooperative \citep{isecbenchmarks} and results from an investigation of nature-based solutions \citep{lunghinoEtAl2019} in Section 2.4, and consider the propagation of a computed tsunami over the observed topography of Crescent City in Section 2.5. 
To quantify the accuracy of the solutions, we test our numerical solver against some classical analytical solutions to the shallow water equations. We assess the model's ability to capture key physical processes relevant to inundation, including steep wave propagation, friction, and topography dependence. We use relative errors in the $L_\infty$ and $L_2$ sense as the metric to determine model accuracy. These are approximated in this paper in the following manner:
\begin{equation}
    \text{$L_\infty$} = \frac{\max_\Omega|h_c - h_a|}{\max_\Omega|h_a|},\quad \text{$L_2$} = \sqrt{\frac{\sum_\Omega(h_c-h_a)^2\Delta\Omega}{\sum_\Omega(h_a)^2\Delta\Omega}} \, ,
\end{equation}
where $h_c$ is the computed solution at the discretized cells, $h_a$ is the analytical solution at the corresponding cells, and $\Omega$ denotes the computational domain.

\subsection{Wet dam break}
\begin{figure}[h]
    \centering
    \includegraphics[width=\textwidth]{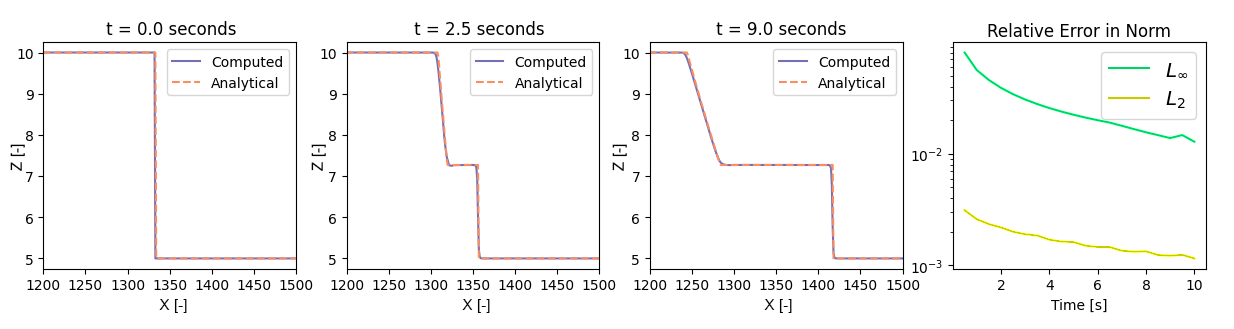}
    \caption{On the left, several instances in time of the computed (purple) water heights to wet dam break compared with the analytical (orange, dashed) water heights. The rightmost figure plots the $L_2$ and $L_{\infty}$ relative norms of the error between the analytical and computed solutions.}
    \label{fig:dam_break}
\end{figure}
The classical one-dimensional Wet Dam Break \citep{stoker_1957} provides us an opportunity to test the ability of our code to capture shock propagation and advection. In this case, there is no friction ($n = 0$) and the topography is flat ($b(x) = 0$). The boundaries are set at a constant height with zero flux. We impose the following initial condition:
\begin{equation}
(hu) = 0, (hv) = 0, h(x) = \begin{cases}h_l & x\leq x_0\\
      h_r & x > x_0 \end{cases} \, ,
\end{equation}
where $h_l$ and $h_r$ are the constant water heights on either side of a shock front $x_0$. We compare our numerical solution for water height against the dynamic analytical solution from \citet{delestre2013swashes}:
\begin{align}
h(x,t)&=\begin{cases} h_l &x \leq x_1\\
                \frac{4}{9g}\left(\sqrt{gh_l}-\frac{x-x_0}{2t}\right)^2 &  x_1(t)<x\leq x_2(t)\\
                \frac{c_m^2}{g} &x_2(t) < x \leq x_3(t)\\
                h_r & x > x_3(t)
\end{cases}\, ,\\
x_1(t) &= x_0 - t\sqrt{gh_l} \, ,\\
x_2(t) &= x_0+t(2\sqrt{gh_l}-3c_m) \, , \\
x_3(t) &= x_0+t\frac{2c_m^2(\sqrt{gh_l}-c_m)}{c_m^2-gh_r} \, , \quad \text{and}\\
c_m &\text{is the solution to} -8gh_rc_m^2(\sqrt{gh_l}-c_m)^2+(c_m^2-gh_r)^2(c_m^2+gh_r)=0 \, .
\end{align}
A qualitative comparison of the computed and analytical solutions for times $t =$0, 2.5, and 9 seconds is shown in the left plots of Fig.~\ref{fig:dam_break}. The relative error between the analytical and computed solutions in the infinity and 2-norms at a small distance away from the shock front are plotted on the right. We interpret the converging relative error norms to a low magnitude as verification of our implementation to sufficiently capture shock propagation and advection.

\subsection{Planar parabolic bowl}
\begin{figure}[h]
    \centering
    \includegraphics[width=\textwidth]{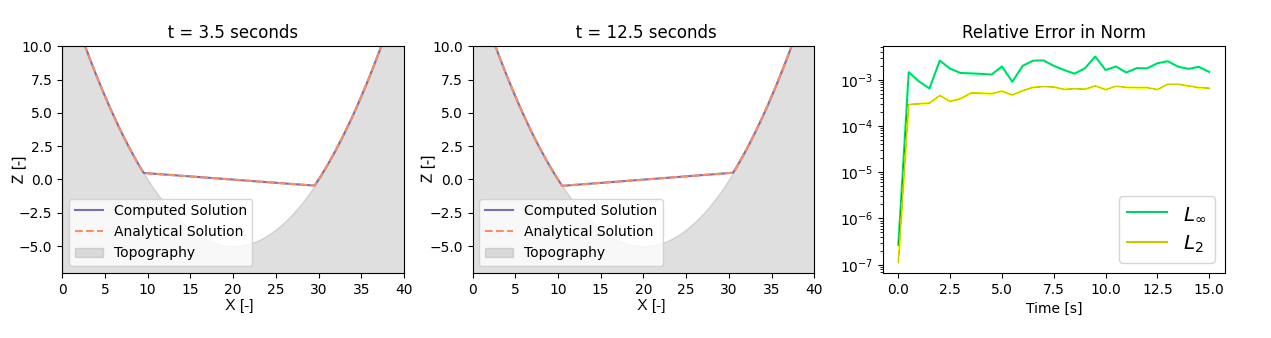}
    \caption{On the left, several instances in time of the computed (purple) water heights to the one-dimensional parabolic bowl compared with the analytical (orange, dashed) water heights. The rightmost figure plots the $L_2$ and $L_{\infty}$ relative norms of the error between the analytical and computed solutions.}
    \label{fig:parabola}
\end{figure}
The classical one-dimensional planar parabolic bowl originally suggested by \citep{Thacker1981}, is an oscillating solution allowing us to test the source term for topography without friction ($n = 0$). We enforce homogeneous Dirichlet conditions in both flux and water height, at a resolution of 1~m. Once again, we take the test directly from \citet{delestre2013swashes}, resulting in the following description of the base topography:
\begin{equation}
    b(x) = h_0\left(\frac{1}{a^2}\left(x-\frac{L}{2}\right)^2-1\right)\, ,
\end{equation}
corresponding with the following initial condition:
\begin{equation}
    (hu) = 0, (hv) = 0, h(x) = \begin{cases} -h_0\left(\left(\frac{2x-L+1}{2a}\right)^2-1\right) & \frac{1-2a+L}{2} < x < \frac{1+2a+L}{2}\\
    0 & \text{otherwise} \end{cases} \, .
\end{equation}
This leads to the following dynamic analytical solution for the water height:
\begin{equation}
    h(x,t) = \begin{cases} -h_0\left(\left(\frac{2x-L}{2a}+\frac{1}{2a}\cos\left(\frac{\sqrt{2gh_0}t}{a}\right)\right)^2-1\right) & x_1(t) < x < x_1(t) + 2a\\
    0 & \text{otherwise} \end{cases} \, ,
\end{equation}
where $x_1(t) = \frac{1}{2}\cos\left(\frac{\sqrt{2gh_0}t}{a}\right)-a+\frac{L}{2}$. A qualitative comparison of the parabolic bowl solution at the time instances $t = $3.5 sand $t = $12.5 s can be seen on the left of Fig.~\ref{fig:parabola}. The analytical and computed solutions appear to correspond to one another well. For a more quantitative analysis,the relative error-norms of the solutions are depicted on the right of Fig.~\ref{fig:parabola}. We interpret the converging relative error norms to a low magnitude as verification of our implementation to sufficiently capture the source term of the shallow water equations induced by topography.

\subsection{Steady flow down a slope with friction}
\begin{figure}[h]
    \centering
    \includegraphics[width=\textwidth]{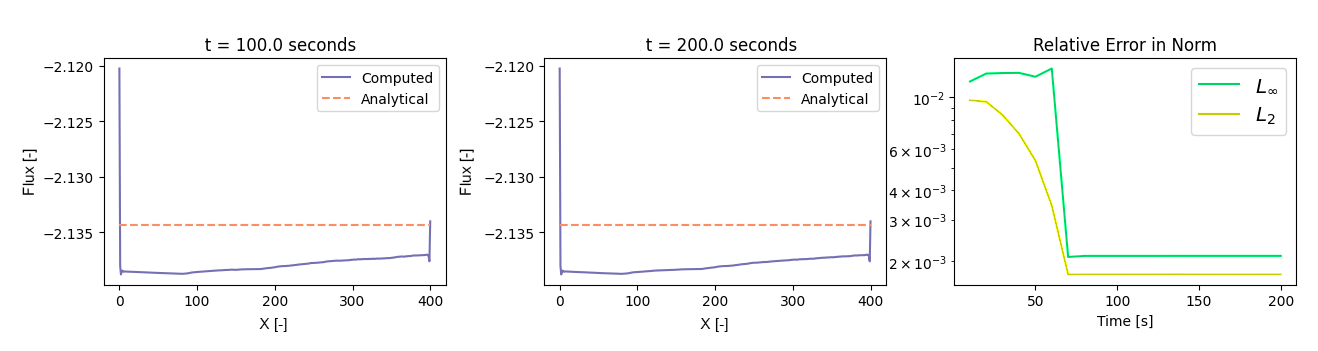}
    \caption{On the left, several instances in time of the computed (purple) fluxes given a water level and a slope, compared to the analytical (orange, dashed) flux. The rightmost figure plots the $L_2$ and $L_{\infty}$ relative norms of the error between the analytical and computed solutions.}
    \label{fig:steady_state}
\end{figure}
We do a short test in order to assess the correctness the discretized friction source term, focusing on a relatively simple flow down a slope with finite friction ($n = 0.033$) as tested by \citet{Xia2018}. The steady state flow down a slope then becomes
\begin{equation}
    (hu) = \frac{\sqrt{b_x}}{n}h^\frac{5}{3}
\end{equation}
where $b_x$ is the slope. In this test, we initialize the problem close to the steady state solution for a wave height of 0.5 and a slope of $\frac{1}{20}$. This specific example and its convergence toward steady state is shown in Fig.~\ref{fig:steady_state}. The left plots show the flux at $t=100~$s and $t = 200~$s. The results are almost identical, indicating that a steady state has been reached. On the right plot, the error norm of the steady state flux takes some time to reach steady state, but reaches a very small level. Because we approach the appropriate steady state solution and achieve a very small error norm, our implementation is verified in capturing a manning friction law.

\subsection{Validation for tsunami simulations}
\begin{figure}[h]
    \centering
    \includegraphics[width=\textwidth]{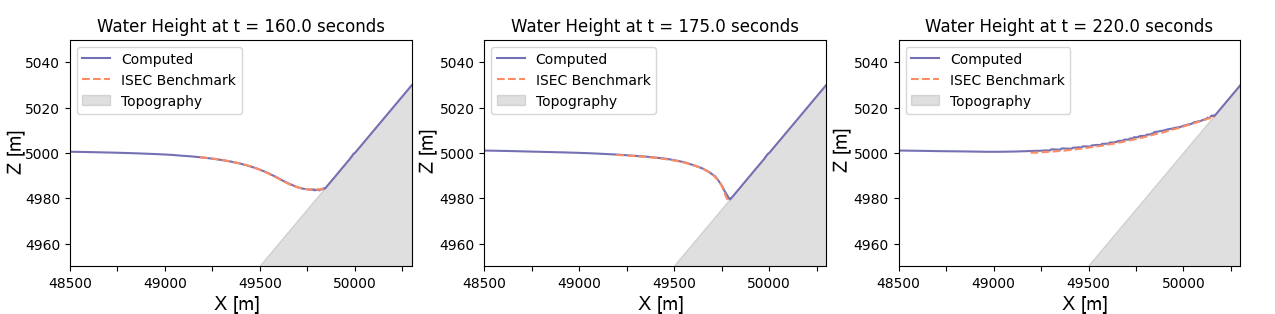}
    \caption{Qualitative comparison of the computed solution with resolution 1~m compared to the ISEC benchmark at three different time instances.}
    \label{fig:isec_benchmark}
\end{figure}

To assess the ability of the code to capture tsunami propagation, we start with a popular numerical benchmark from the Inundation and Science Engineering Cooperative (ISEC) \citep{isecbenchmarks} that represents tsunami runup over an idealized planar beach that provides solutions for tsunami runup at times $t 
= $, 180 s, 195 s, 220 s. We formulate the initial condition for water height using \citet{lunghinoEtAl2019}. The solutions from the benchmark (dashed, orange) are qualitatively compared with the numerical solution produced by our code (solid, purple) in Fig.~\ref{fig:isec_benchmark}. We take the qualitative agreement as validation of the model's ability to model the runup of a Carrier N-Wave \citep{carrier2003tsunami}. 

\begin{figure}[h]
    \centering
    \includegraphics[width=\textwidth]{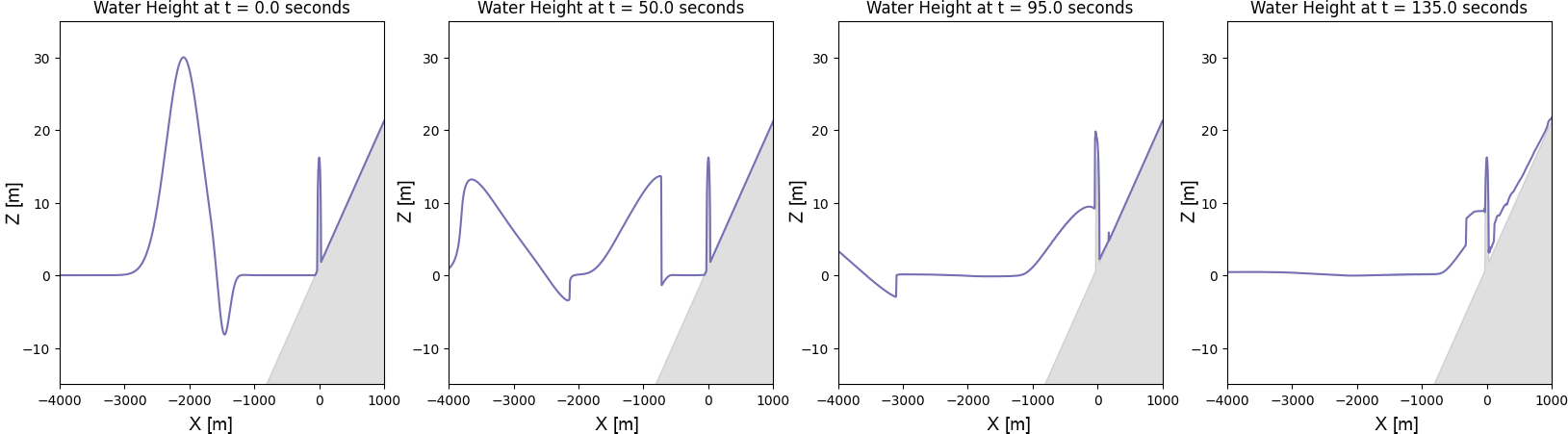}
    \caption{Several snapshots in time of the tsunami's propagation over a modeled ellipsoidal hill on a slope. From left to right, the formulation of the initial Carrier N-wave at $t = 0$, followed by the propagation of a wave front toward the hill at $t = 50$, collision of the wave front of the hill at $t = 95$, and the formation of a reflected wave at $t = 135$.}
    \label{fig:tsunami_with_hill}
\end{figure}

Since we are interested in leveraging TPUs for tsunami-risk mitigation planning, we take a look at the ability of our shallow water equation code to reproduce a few particular results by \citet{lunghinoEtAl2019} who investigated the effects of hills on a tsunami running up on a planar beach. The tsunami is initialized as Carrier's N-wave \citep{carrier2003tsunami}:
\begin{equation}
\label{carrierWaveEQN}
\eta = 2 ( a_1 \exp\{ -\hat{k}_1(x - \hat{x}_1)^2\} - a_2\exp\{ \hat{k}_2(x - \hat{x}_2)^2 \} ) ,
\end{equation}
where $\eta = h + z$, $\hat{x}_1 = 1000 + 0.5151125 \lambda $, $\hat{x}_2 = 1000 + 0.2048\lambda$, $\hat{k}_1 = 28.416/ \lambda^2$, $\hat{k}_2 = 256 /\lambda^2$, $a_1 = A$, and $a_2 = A/3$. While this is the analytically correct form, the flow origin in the code is not the shoreline,  so there are some effective shifts $\hat{x}_1$ and $\hat{x}_2$ that we need to do. An example of the Carrier wave initial condition and offshore propagation behavior for $A=15$ and $\lambda=2000$ is shown in Figure \ref{fig:tsunami_with_hill}. We apply free slip, no-penetration boundary conditions to the four domain boundaries, which means that that the component of the boundary-normal component of the velocity vector is zero whereas its tangential component is unaltered.
The shallow water equation model presented in this study is able to reproduce the wave reflection provided by a hill, consistent with results from \citet{lunghinoEtAl2019}. Because this simulation is possible by the implementation, other further analysis can be conducted to understand the mitigative benefit of other nature-based solutions. 

\subsection{Real-world scenario}
\begin{figure}[h]
    \centering
    \includegraphics[width=\textwidth]{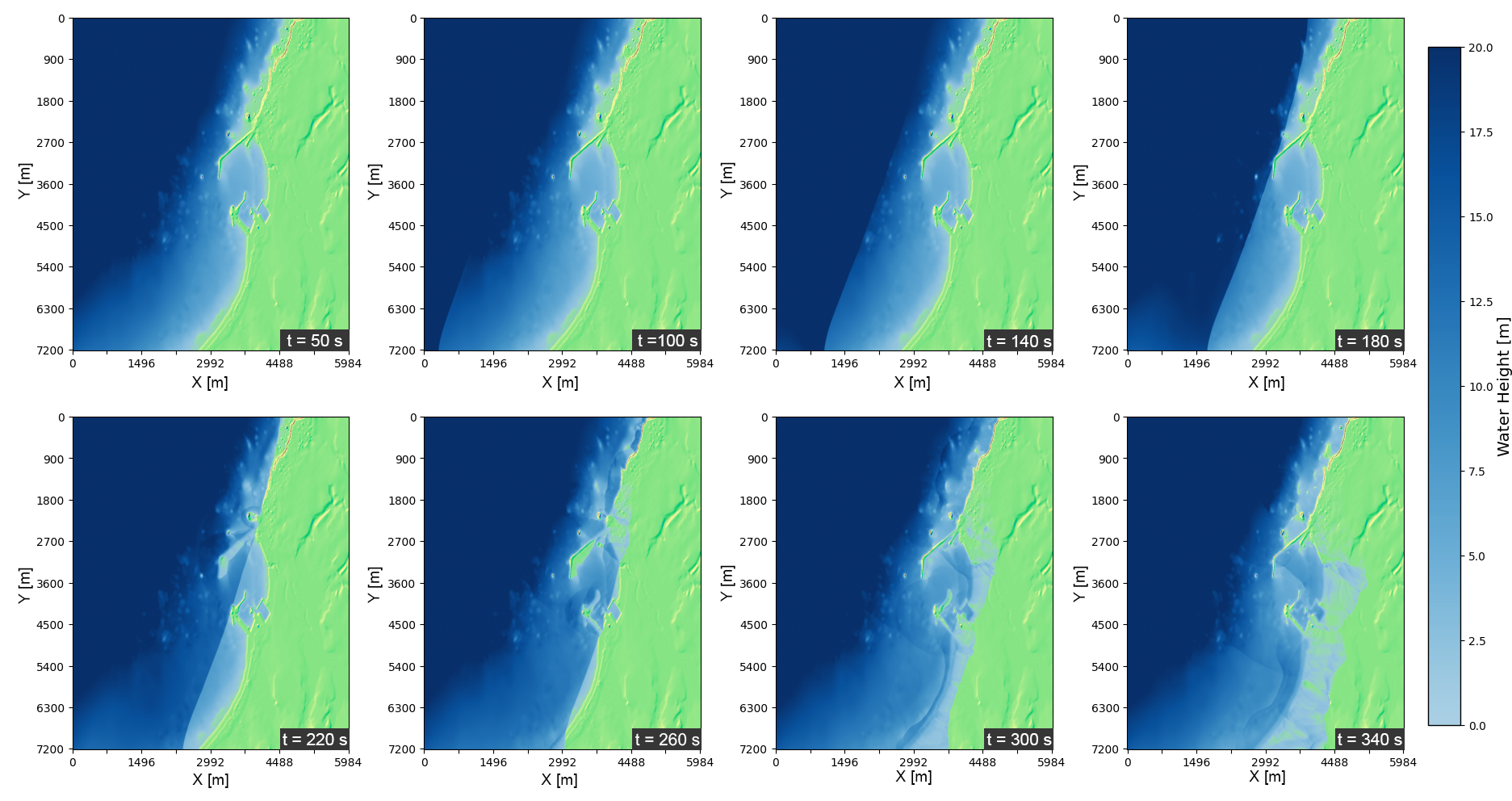}
    \caption{Several snapshots of modeled tsunami propagation over terrain and geological features of Crescent City, CA, where any level of blue indicates water cover and green depicts a stylized map of the topography above surface level. From left to right, then top to bottom, we have steady state near shore at $t = 50$; followed by the propagation of a wave front at $t = 100$ and $140$; contact with Crescent City harbor at $t = 180$; inundation of the harbor and some of the coastline at $t = 220$ and $260$; and tsunami reflection and inundation at $t = 300$ and $340$.}
    \label{fig:cresent_city}
\end{figure}

Past tsunamis impacting the West Coast of the United States have caused more damage around the harbor of Crescent City in California than elsewhere along the Pacific Coast \citep{arcasUslu2010noaa}. For this reason, we chose an area of approximately 105 ${\rm km^2}$ around Crescent City to demonstrate the code's ability to capture the impact of an idealized tsunami event for a real location at high resolution. To approximate the actual bathymetry and topography, we use a Digital Elevation Model for this area with uniform grid spacing of 4~m provided by NOAA \citep{crescentcitydem,crescentcitydemreport}. For the sake of providing a proof of concept, we initialize the tsunami with the idealized waveform described above, with slightly adjusted parameters: $A=10$, $\lambda=2000$, $\hat{x}_1 = 6000 + 0.5151125 \lambda $, and $\hat{x}_2 = 6000 + 0.2048\lambda$. The chosen parameters lead to maximum inundation patterns similar to that seen in one modeled extreme scenario from \citet{arcasUslu2010noaa}. In Fig.~\ref{fig:cresent_city}, we start with an absence of any nearshore wave (including at $t = 50 s$) and then a development of a tsunami front that is visible to the shoreline by $t = 140 s$. That front penetrates the harbor by $t = 220 s$, and is soon followed by the inundation of the coastline as well as reflection of wave energy back to the ocean. We also observe that the mountain range on the upper part of the figure clearly provides a significant protective benefit to the land beyond it.

%

\section{Performance analysis on TPU}
\subsection{Number of TPU cores}
In communities where users may not have access to high performance computing facilities, the Cloud TPU Platform provides a unique ability to perform large-scale computations, and perform them rapidly. To address this potential benefit, we first measure the wall-clock time of a simulation of a tsunami reaching Crescent City using a varying number of TPU cores on one TPU device. As shown in Table~\ref{table:num_cores}, the problem size posed by the realistic scenario is sufficient to see rapid improvements in runtime based on the number of cores.
\begin{table}[h!]
    \centering
    \begin{tabular}{|l|l|l|l|l|}
    \hline
    Number of Cores & 1 & 2 & 4 & 8 \\
    \hline
    TPU Runtime [s]    & 6894 & 3876 & 2000 & 1036\\
    \hline
    \end{tabular}
    \vspace{5 mm}
    \caption{Approximate TPU Runtimes (in seconds) with varying numbers of TPU cores for Crescent City Configuration using time step of $\Delta t = 5\cdot 10^{-3}$, Total Array Size of approximately 1802 by 3984 elements (4 meter resolution), cores all aligned in the y-direction as suggested in \citep{pierceEtAl2022}. This runtime excludes transfer times between the CPU and TPU.}
    \label{table:num_cores}
\end{table}

\subsection{Geophysical problem resolution}
\begin{figure}[H]
    \centering
    \includegraphics[width=0.5\textwidth]{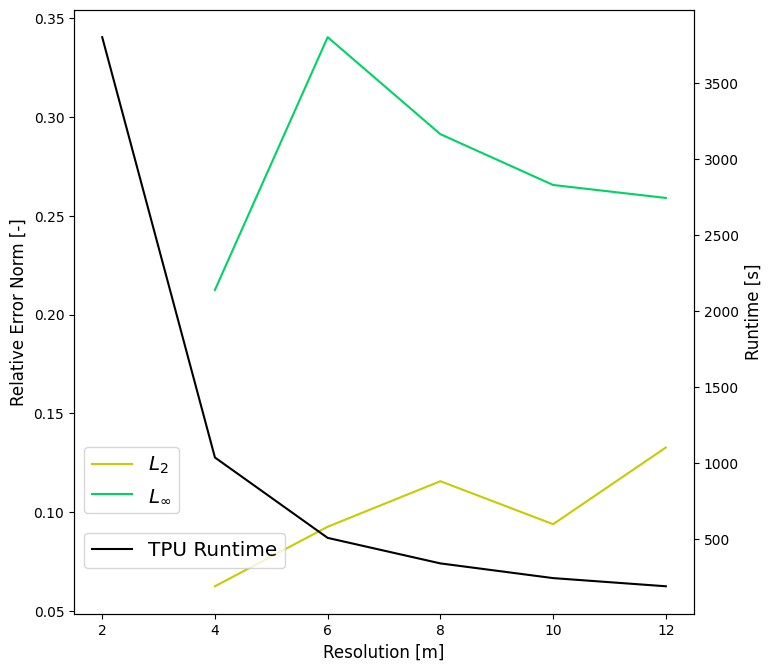}
    \caption{Graphical depiction of the TPU runtimes (purple) and relative error norms (blue) under varying resolutions for computing the tsunami propagation over the Crescent City DEM.}
    \label{fig:cresent_city_resolutions}
\end{figure}

The typical realistic scenario that decision-makers will face will involve large problem sizes due to both the extent of their spatial domain, but also the level of resolution necessary to model tsunami propagation and inundation over complex topography. Therefore, we check convergence and runtime under varying degrees of resolution for the current realistic scenario as well, depicted graphically in \ref{fig:cresent_city_resolutions} and in table \ref{table:cresent_city_resolutions}.
\begin{table}[h!]
    \centering
    \begin{tabular}{|l|l|l|l|l|l|l|}
    \hline
Resolution [m]        & 2        & 4        & 6        & 8        & 10       & 12       \\ \hline
Runtime [s]         & 3805.167 & 1035.719 & 506.0652 & 337.9506 & 241.0804 & 187.3016 \\ \hline
Number of Elements & 28713068 & 7179168  & 3192512  & 1794792  & 1149274  & 798128   \\ \hline
Efficiency         & 0.000133 & 0.000144 & 0.000159 & 0.000188 & 0.00021  & 0.000235 \\ \hline
$L_2$ Error Norm           & *        & 0.062475 & 0.092618 & 0.115671 & 0.093901 & 0.132735 \\ \hline
$L_{\infty}$ Error Norm         & *        & 0.212536 & 0.340507 & 0.291389 & 0.265669 & 0.259101 \\ \hline
    \end{tabular}
    \vspace{5 mm}
    \caption{Approximate TPU Runtimes (in seconds) with varying resolutions for Crescent City Configuration using time step of $\Delta t = 5\cdot 10^{-3}$, using the 2~m resolution as a benchmark for correctness. Ran on a single TPU with 8 cores.}
    \label{table:cresent_city_resolutions}
\end{table}

\begin{figure}[H]
    \centering
    \includegraphics[width=\textwidth]{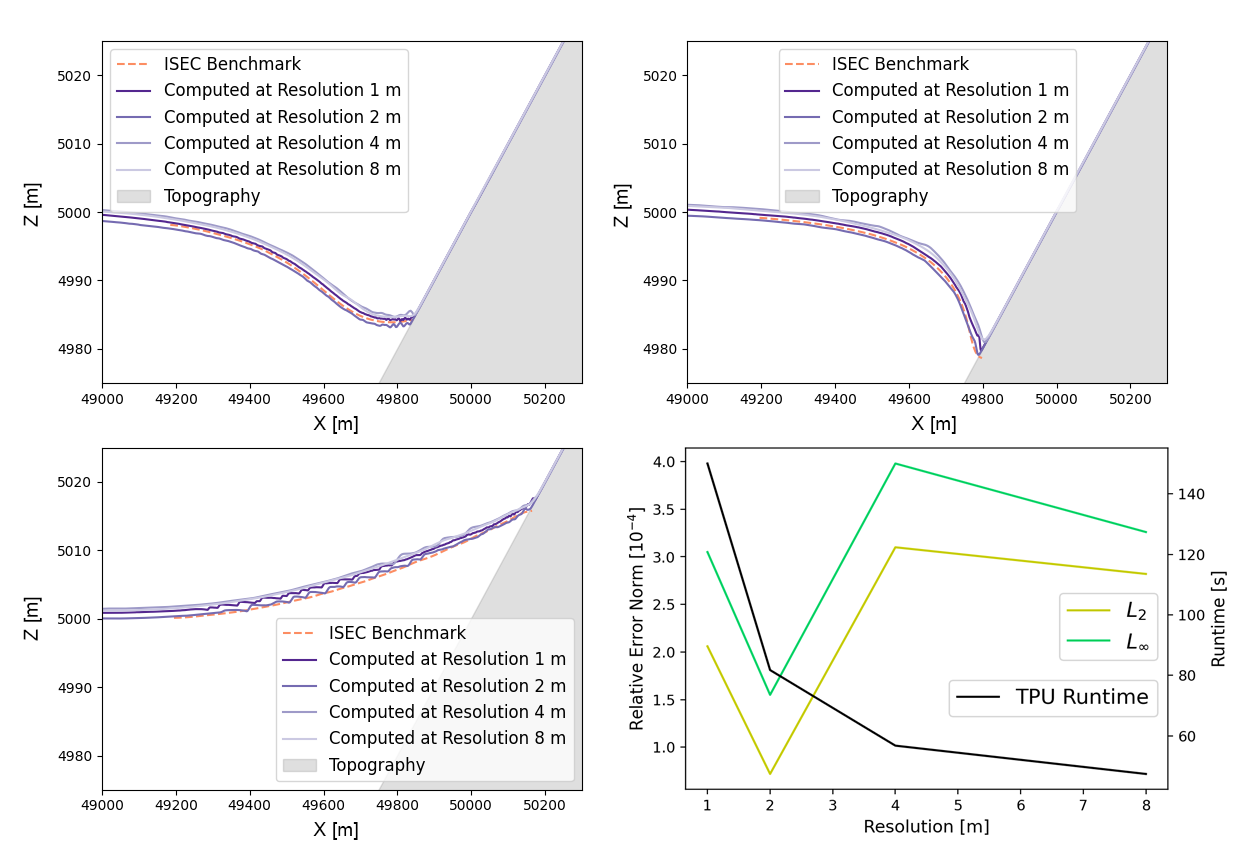}
    \caption{Graphical depiction of the TPU runtimes (purple) and relative error norms (blue) under varying resolutions for computing the tsunami propagation in the ISEC Benchmark.}
    \label{fig:tsunami_resolutions}
\end{figure}

We finally perform the same analysis under varying degrees of resolution using the benchmark from the Inundation Science and Engineering Cooperative \citep{isecbenchmarks} that we previously validated in Fig.~\ref{fig:isec_benchmark}. A qualitative comparison of the tsunami propagation under different resolutions are graphically depicted in Fig.~\ref{fig:tsunami_resolutions} in the top two and bottom left figure. In the bottom right plot of Fig.~\ref{fig:tsunami_resolutions},  we see the expected fall in runtime based on coarser resolution (purple), and a rise in relative error (with the exception of the highest resolution 1~m). The corresponding values are documented in \ref{table:tsunami_resolutions}.

\begin{table}[h!]
    \centering
    \begin{tabular}{|l|l|l|l|l|}
    \hline
    Resolution [m]          & 1        & 2        & 4        & 8    \\ \hline
    Runtime [s]    & 150.0193 & 81.74382 & 56.74928 & 47.3934  \\ \hline
    Number of Elements & 1060521  & 277761   & 75756    & 18939   \\ \hline
    Efficiency         & 0.000141 & 0.000294 & 0.000749 & 0.002502  \\ \hline
    $L_2$ Error           & 0.000206 & 7.19E-05 & 0.00031  & 0.000282 \\ \hline
    $L_{\infty}$ Error Norm         & 0.000305 & 0.000155 & 0.000398 & 0.000326  \\ \hline
    \end{tabular}
    \vspace{5 mm}
    \caption{Approximate TPU Runtimes (in seconds) with varying resolutions for the ISEC Tsunami Benchmark using time step of $\Delta t = 5\cdot 10^{-3}$. Ran on a single TPU with 8 cores.}
    \label{table:tsunami_resolutions}
\end{table}

\subsection{Comparison with GeoClaw}
\begin{figure}[H]
    \centering
    \includegraphics[width=\textwidth]{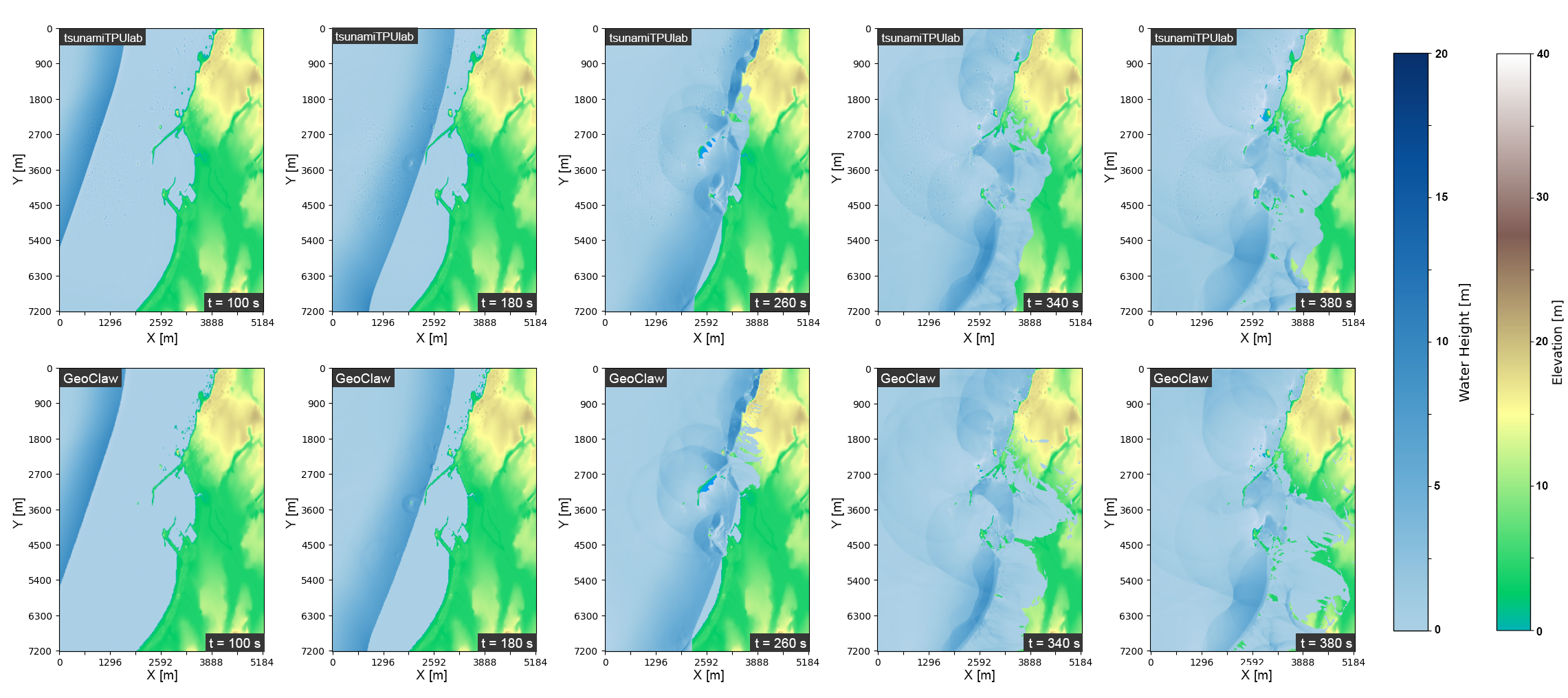}
    \caption{TPU solution (top row) at several time instances compared to the GeoClaw solution (bottom row). The arrival of the tsunami front ($t = 100, 180$), the inundation of the harbor ($t = 260$), and coastal inundation and reflection is depicted, and relatively comparable.}
    \label{fig:tsunami_cpu_comparison}
\end{figure}
We used GeoClaw \citep{clawpack,mandli2016clawpack,berger2011geoclaw} ran on a single thread of a CPU (Intel i7-8650 with a base frequency of 1.9 GHz) to compare numerical solutions to a tsunami propagation in order to assess performance enhancements provided by the TPU. While our TPU-based code completes the 400 second simulation in approximately 17 minutes of wall-clock time, the GeoClaw implementation on the CPU takes approximately 630 minutes. Comparisons of the two numerical solutions can be seen in Fig.~\ref{fig:tsunami_cpu_comparison}, where the top row includes several instances in time of the TPU numerical solution, and the bottom row depicts the GeoClaw numerical solution at the same instances in time. Although some differences can be seen in the geometry of inundation by $t = 380 s$ in the rightmost plots, the solutions do generally appear similar over time, lending credibility to the validity of the numerical solution.

\subsection{Energy utilization}
Estimates of energy efficiency of computing operations are becoming increasingly popular, especially in response to progressing climate change \citep{fuhrerEtAl2018, fouresteyEtAl2014}. To address this potential benefit provided by TPUs, we conduct a heuristic analysis of the energy savings of running these operations on a TPU rather than a CPU. We base an order-of-magnitude estimate on the claims of Google in the maximum power efficiency of the TPU, 2 trillion operations per second (TOPS) per Watt \citep{googleTPU}. We use this heuristic for power efficiency, consider the tsunami propagation problem under a 8~m resolution for Crescent City, and manually approximate the number of floating point operations that our implementation performs in each quadrature step at this resolution. We estimate that the TPU performs approximately 11.9 million floating-point operations for each simulated time step, and we see that each time-step takes an average of about 4.2 milliseconds, corresponding to about 2.83 TOPS. Assuming constant power efficiency regardless of capacity, this translates to 1.41 Watts and an energy usage of 1.18 J for each simulated second given our current time-step configuration. Combining with the total runtime for a 400 modeled-second simulation of Crescent City, we see a total cost of approximately $4.76\times  10^{2}$ J = $1.32\times  10^{-1}$ Wh of energy. At a price of 21 cents/kWh in the U.S. at the time of writing this article, this simulation has a monetary cost of $2.7\times  10^{-3}$ cents.

We use the aforementioned GeoClaw run as our CPU comparison on energy utilization in order to get an order of magnitude estimate of the power savings. The particular processor for this CPU comparison, the Intel i7-8650 with a base frequency of 1.9 GHz, has a Thermal Design Power of 15 W for 8 total threads. Each simulated second took approximately 96.4 seconds of CPU runtime, which translates to approximately 181 Joules for each simulated second of a grid with 8~m resolution. Given the order of magnitude estimates, we note that the 400 modeled-second simulation would see a total cost of approximately $7.24\times  10^{4}$ J = $2.01\times  10^{1}$ Wh of energy, or a monetary cost of 0.04 cents, a cost multiplicative factor of 20. We linearly extrapolate to estimate the number of CPU threads needed to match the runtime speed of our model using the TPU, and find that approximately 37 CPU threads would be needed. With this in mind, we find that the cost multiplicative factor for a CPU simulation of performance equivalent to that of a TPU would be closer to an order of 700. TPU energy savings for high performance are clearly substantial, and not ignorable.

\section{Discussion}
Sustainable tsunami-risk mitigation in the Pacific Northwest is a challenging task. Some challenges come from beneath, because previous large subduction zone earthquakes at Cascadia led to $0.5-1$ m of co-seismic subsidence, the sudden sinking of land during an earthquake \citep{wang2013heterogeneous}. Strong shaking can also lead to liquefaction \citep{atwater1992geologic,takada2004evidence}. Other challenges come from the ocean, where sea-level rise \citep{church200620th,bindoff2007observations} and intensifying winter storms \citep{graham2001evidence} have increased wave heights \citep{ruggiero2010increasing, ruggiero2013intensifying} and accelerated coastal erosion \citep{ruggiero2008impacts}. A recent USGS report documented rapid shoreline changes at an average rate of almost 1 m/yr across 9,087 individual transects \citep{ruggiero2013national}, suggesting the possibility that the shoreline might change significantly during the century-long return-period of large earthquakes in Cascadia \citep{witter2003great}.

The picture that emerges is that of a highly dynamic coastline -- maybe too dynamic for an entirely static approach. Nature is not only continuing to shape the coastline, but is also a fundamental component of the region's cultural heritage, identity and local economy. So, it is maybe not surprising that the Pacific Northwest is a thought-leader when it comes to designing hybrid approaches to sustainable climate adaptation through the Green Shores program \citep{dalton2013climate} and to vertical tsunami evacuation through Project Safe Haven \citep{safeHavenV1}. 

Project Safe Havens is a grass-roots approach to reducing tsunami risk mostly by providing accessible vertical-evacuation options for communities. Many proposed designs entail reinforced hillscapes like the one shown in figure 2, intended to dissipate wave energy and provide vertical evacuation space during tsunami inundation. To build confidence in such a solution and its mitigation effects, risk managers must be able to quickly and precisely forecast a tsunami inundation, preferably via a publicly available, centralized modeling infrastructure.

This paper is meant to be a first step towards a major community based infrastructure that will allow local authorities around the world to readily execute tsunami simulation once a tsunami in their proximity has been detected. We aim to provide a proof-of-concept rather than a complete implementation. As such, we used a very similar base framework used by \cite{pierceEtAl2022} of halo exchange in combination with a WENO and Runge-Kutta scheme, which may not be optimally taking advantage of the TPU's computing structure and capabilities. We originally chose these schemes to maintain higher order accuracy and ease of implementation but, eventually, a convolution-based implementation of the quadrature of the shallow water equations should be tested for maximum performance utilization of the TPUs.

Because our code is specifically an implementation of the shallow water equations for the TPU, it is currently unable to model tsunami initiation, or any fluid structure interactions that may be desired to accompany analysis of nature-based solutions. Instead, it requires an initial condition for wave heights and fluxes, meaning a full tsunami simulation would require coupling the results of a tsunami initiation model as an input. While our implementation is relatively limited in scope, the model is certainly able to provide a starting point for a more complete software package for communities as they evaluate nature-based options for tsunami mitigation.

Though just a starting point for a remotely-available package, we successfully replicated some of the results found about coastal mitigation parks posed by \cite{lunghinoEtAl2019}, and we were able to model tsunami runup over with the real bathymetry around Crescent City, in California, included in the code by means of a DEM file from the USGS digital elevation database; for comparison purposes, we ran the same test using the popular open-source solver GeoClaw \citep{clawpack,berger2011geoclaw}. The results of the two models are in good agreement, as shown in Fig.~\ref{fig:tsunami_cpu_comparison}. Our contribution lies in demonstrating an enhanced ability to run high quality simulations using the TPU available remotely using Google Cloud Platform. As a result, high quality tsunami simulations are available to remote communities for rapidly evaluating mitigation or evacuation options when faced with coastal flooding.

We argue that TPUs are preferable to large, heavily parallel simulations on CPU or GPUs, because the TPU-based simulations we show here do not require access to large computing clusters. These are usually made available to scientists and engineers by supercomputing centers around the world by means of competitive grants for computing time or by use of the cloud offered by private companies. However, an expert user knowledge of these systems from a scientific computing perspective is necessary to design, run, and interpret model results, and the compute infrastructure itself may not be available to early warning centers in many parts of the world. In contrast, our code is available on Github and fully implemented in Python, can be ran through a web browser, and visualized through a simple notebook file using Google Colab. While performance can be enhanced with some knowledge about TPU architectures, community risk managers do not need this knowledge to run high quality tsunami simulations rapidly for real, physical domains with associated DEMs.

Finally, in the face of rising energy costs in both a monetary and climate sense, the TPU infrastructure allows the support of more energy-efficient simulations over those of CPU-based clusters. This means that those coastal flood risk managers in remote communities immediately are able to support design decisions with model results in a climate-friendly manner.

Although not our focus here, we note our approach may also contribute to early tsunami warning. Once triggered, tsunamis move fast; this fact makes it necessary to model and assess their potential for damage ahead of time once they have been detected offshore. For a sufficiently fast early warning and prompt evacuation, the tsunami modeling infrastructure has an important time constraint \citep{gilesEtAl2021} to be considered, and Faster Than Real Time (FTRT) simulations are necessary \citep{behrens2021probabilistic, lovholtEtAl2019}. To make FTRT simulations a reality, tsunami models are being rewritten or adapted to run on Graphical Processing Units (GPUs) \citep{lovholtEtAl2019, behrensDias2015, satria2012}. A TPU-based implementation as proposed here might be another meaningful step into that direction.

\section*{Author contributions} \textbf{Ian Madden}: Software, Analysis, Writing. \textbf{Simone Marras, PI}: Conceptualization, Methodology, Writing, Supervision. \textbf{Jenny Suckale}: Conceptualization, Methodology, Writing, Supervision. 

\section*{Competing interests}
The authors declare that they have no conflict of interest.

\section*{Data availability statement}
Our work is available as a GitHub release at \url{https://github.com/smarras79/tsunamiTPUlab/releases/tag/v1.0.0} or on archive at \url{10.5281/zenodo.7574655}.

\section*{Acknowledgements}
This work was supported by the National Science Foundation's Graduate Research Fellowships Program (GRFP) awarded to Ian Madden.

\bibliography{bibliography_completa}

\section{Appendix}
\subsection{Running the code}
Due to the restrictions of the TPU using Google Cloud Storage, Google's buckets will need to be used in order to run the notebooks. With a computing project setup on Google Cloud, users can quickly run any of the example notebooks or design their own simulation. Any of the example notebooks available on GitHub (with the exclusion of \texttt{tpu\_tsunami.ipynb}, which contains the full implementation with all of the different scenarios; and \texttt{Create\_Scenarios.ipynb}, which can aid users in generating a custom DEM file) can be quickly ran by going through the notebook after a few early setup steps. 
\begin{enumerate}
\item Download the TPU-Tsunami Repository from \url{https://github.com/smarras79/tsunamiTPUlab/releases/tag/v1.0.0} to your local machine. Create a project on Google Cloud Platform and associate a publicly available bucket with the project.
\item Modify the \texttt{user\_constants.py} file to specify the \texttt{PROJECT\_ID} and \texttt{BUCKET} with the specifics of your Google Cloud Project. If you wish to change some simulation constants, modify the beginning of the \texttt{tpu\_simulation\_utilities.py} file.
\item Navigate to \url{https://colab.research.google.com/} and open the example notebook (or your own notebook) from the TPU-Tsunami Repository using Colab's open from Github tool.
\item Navigate to Runtime > Change runtime type, and verify that the TPU option is chosen as the Hardware Accelerator.
\item Upload your \texttt{user\_constants.py} and \texttt{tpu\_simulation\_utilities.py} files to your notebook session using the drag-and-drop feature under Files. Upload any corresponding DEM files to the session as well.
\item Specify a function corresponding to an initial condition for your DEM file (or use one example initial condition).
\item Set initial conditions, boundary conditions as clarified in the bottom of any example notebook run. Set last simulation parameters defining numerical resolution (\texttt{resolution}), time step size (\texttt{dt}), output file times, TPU core configuration (currently only capable of variation of \texttt{cy}), and DEM file name on bucket (\texttt{dem\_bucket\_filename}).
\item Run the simulation.
\item Analyze results. 
\end{enumerate}

\end{document}